\documentclass[sn-standardnature,iicol]{sn-jnl}

\usepackage{siunitx}
\usepackage{amssymb}
\usepackage{times}
\usepackage{booktabs}
\usepackage{amsmath, array, multirow, rotating, makecell}

\graphicspath{{./figures/}}

\newcommand\aj{{Astron.~J.}}
\newcommand\apj{{Astrophys.~J.}}
\newcommand\apjl{{Astrophys.~J.~Lett.}}
\newcommand\apjs{{Astrophys.~J.~Suppl.}}

\newcommand\aap{{Astron.~Astrophys.}}
\newcommand\mnras{{Mon.~Not.~R.~Astron.~Soc.}}
\newcommand\nat{{Nature}}
\newcommand\araa{{Ann.~Rev.~Astron.~Astrophys.}}
\newcommand\jqsrt{{J.~Quant.~Spectrosc.~Radiat.~Transf.}}

\newcommand\pasp{{Publ. Astron. Soc. Pac.}}
\newcommand\icarus{{Icarus}}

\jyear{2023}%

\begin{document}

\title{\centering Methane Throughout the Atmosphere of the Warm Exoplanet WASP-80b}

\author[1,2]{\fnm{Taylor J.} \sur{Bell}}

\author[3,4]{\fnm{Luis} \sur{Welbanks}}

\author[5]{\fnm{Everett} \sur{Schlawin}}

\author[3]{\fnm{Michael R.} \sur{Line}}

\author[6]{\fnm{Jonathan J.} \sur{Fortney}}

\author[2]{\fnm{Thomas P.} \sur{Greene}}

\author[6,7]{\fnm{Kazumasa} \sur{Ohno}}

\author[8]{\fnm{Vivien} \sur{Parmentier}}

\author[9]{\fnm{Emily} \sur{Rauscher}}

\author[10]{\fnm{Thomas G.} \sur{Beatty}}

\author[6]{\fnm{Sagnick} \sur{Mukherjee}}

\author[3]{\fnm{Lindsey S.} \sur{Wiser}}

\author[11]{\fnm{Martha L.} \sur{Boyer}}

\author[5]{\fnm{Marcia J.} \sur{Rieke}}

\author[11]{\fnm{John A.} \sur{Stansberry}}


\affil[1]{\orgdiv{Bay Area Environmental Research Institute}, \orgname{NASA's Ames Research Center}, \orgaddress{\city{Moffett Field}, \state{CA}, \country{USA}}}

\affil[2]{\orgdiv{Space Science and Astrobiology Division}, \orgname{NASA's Ames Research Center}, \orgaddress{\city{Moffett Field}, \state{CA}, \country{USA}}}

\affil[3]{\orgdiv{School of Earth and Space Exploration}, \orgname{Arizona State University}, \orgaddress{\city{Tempe}, \state{AZ}, \country{USA}}}
\affil[4]{NHFP Sagan Fellow}

\affil[5]{\orgdiv{Steward Observatory}, \orgname{University of Arizona}, \orgaddress{\city{Tucson}, \state{AZ}, \country{USA}}}

\affil[6]{\orgdiv{Department of Astronomy and Astrophysics}, \orgname{University of California Santa Cruz}, \orgaddress{\city{Santa Cruz}, \state{CA}, \country{USA}}}

\affil[7]{\orgdiv{Division of Science}, \orgname{National Astronomical Observatory of Japan}, \orgaddress{\city{Tokyo}, \country{Japan}}}

\affil[8]{\orgdiv{Laboratoire Lagrange}, \orgname{Observatoire de la Côte d’Azur, Université Côte d’Azur}, \orgaddress{\city{Nice}, \country{France}}}

\affil[9]{\orgdiv{Department of Astronomy}, \orgname{University of Michigan}, \orgaddress{\city{Ann Arbor}, \state{MI}, \country{USA}}}

\affil[10]{\orgdiv{Department of Astronomy}, \orgname{University of Wisconsin-Madison}, \orgaddress{\city{Madison}, \state{WI}, \country{USA}}}

\affil[11]{\orgname{Space Telescope Science Institute}, \orgaddress{\city{Baltimore}, \state{MD}, \country{USA}}}


\abstract{
The abundances of major carbon and oxygen bearing gases in the atmospheres of giant exoplanets provide insights into atmospheric chemistry and planet formation processes\cite{oberg2011co,madhusudhan2012co}. Thermochemistry suggests that methane should be the dominant carbon-bearing species below $\sim$1000\,K over a range of plausible atmospheric compositions\cite{Burrows2001}; this is the case for the Solar System planets\cite{Adel1934} and has been confirmed in the atmospheres of brown dwarfs and self-luminous directly imaged exoplanets\cite{PPIV2022}. However, methane has not yet been definitively detected with space-based spectroscopy in the atmosphere of a transiting exoplanet\cite{stevenson2010,desert2011,benneke2019,triaud2015,swain2008,gibson2011}, but a few detections have been made with ground-based, high-resolution transit spectroscopy\cite{giacobbe2021,guilluy2022} including a tentative detection for WASP-80b\cite{carleo2022}. Here we report transmission and emission spectra spanning 2.4--4.0 micrometers of the 825\,K warm Jupiter WASP-80b taken with JWST's NIRCam instrument, both of which show strong evidence for methane at greater than 6-sigma significance. The derived methane abundances from both viewing geometries are consistent with each other and with solar to sub-solar C/O and ${\sim}5\times$ solar metallicity, which is consistent with theoretical predictions\cite{Kreidberg+14b,Welbanks+19b,Bean+23}.}


\maketitle
\clearpage

\section*{Main}\label{sec:main}

The WASP-80 system is comprised of a bright (K=8.3\,mag) and cool \mbox{K7--M0 V} star orbited by a 0.95\,R$_{\rm Jup}$, 0.54\,M$_{\rm Jup}$ planet on a 3 day orbit\cite{triaud2013}. WASP-80b is one of exceptionally few known giant exoplanets orbiting low-mass stars\cite{bryant2023}, and the planet has one of the highest planet-to-star mass ratios known to date. WASP-80b was chosen as part of the MANATEE NIRCam$+$MIRI GTO program (GTOs 1185 and 1177; ref.\cite{schlawin2018}) in part because the planet's 825\,K equilibrium temperature\cite{triaud2015} places it in an interesting transitional regime where equilibrium chemistry models predict that there should be detectable CH$_4$ \textit{and} CO/CO$_2$ features in the planet's transmission and emission spectra\cite{moses2013}. Warm, giant exoplanets around M-dwarf stars are also perhaps the most likely planets to exhibit detectable methane signatures due to their large radius ratios and the cool temperature of the stars, as well as the lower metallicity of giant planets compared to sub-Neptunes which favors methane over CO.

Previous low-resolution transmission observations of WASP-80b have shown a mostly featureless optical transmission spectrum\cite{fukui2014}
with a potential Rayleigh scattering slope\cite{wong2022}, a weak H$_2$O feature in the near-infrared\cite{tsiaras2018,wong2022}, no detectable methane\cite{triaud2015,wong2022}, and a potential detection of CO or CO$_2$ from \textit{Spitzer}'s 4.5\,$\mu$m photometric band\cite{triaud2015,wong2022}. More recently, ref.\cite{carleo2022} detected methane in the transmission spectrum of WASP-80b using ground-based, high-resolution spectroscopy, but their detection was only at 4.1\,$\sigma$ (0.1\,$\sigma$ above their detection threshold) and they could not obtain abundance constraints. Meanwhile, secondary eclipse (hereafter just `eclipse') observations of WASP-80b from \textit{Spitzer}'s 3.6 and 4.5\,$\mu$m photometric bands were consistent with a simple $851^{+13}_{-14}$\,K blackbody\cite{triaud2015,wong2022}.

\begin{figure*}
    \centering
    \includegraphics[width=\linewidth]{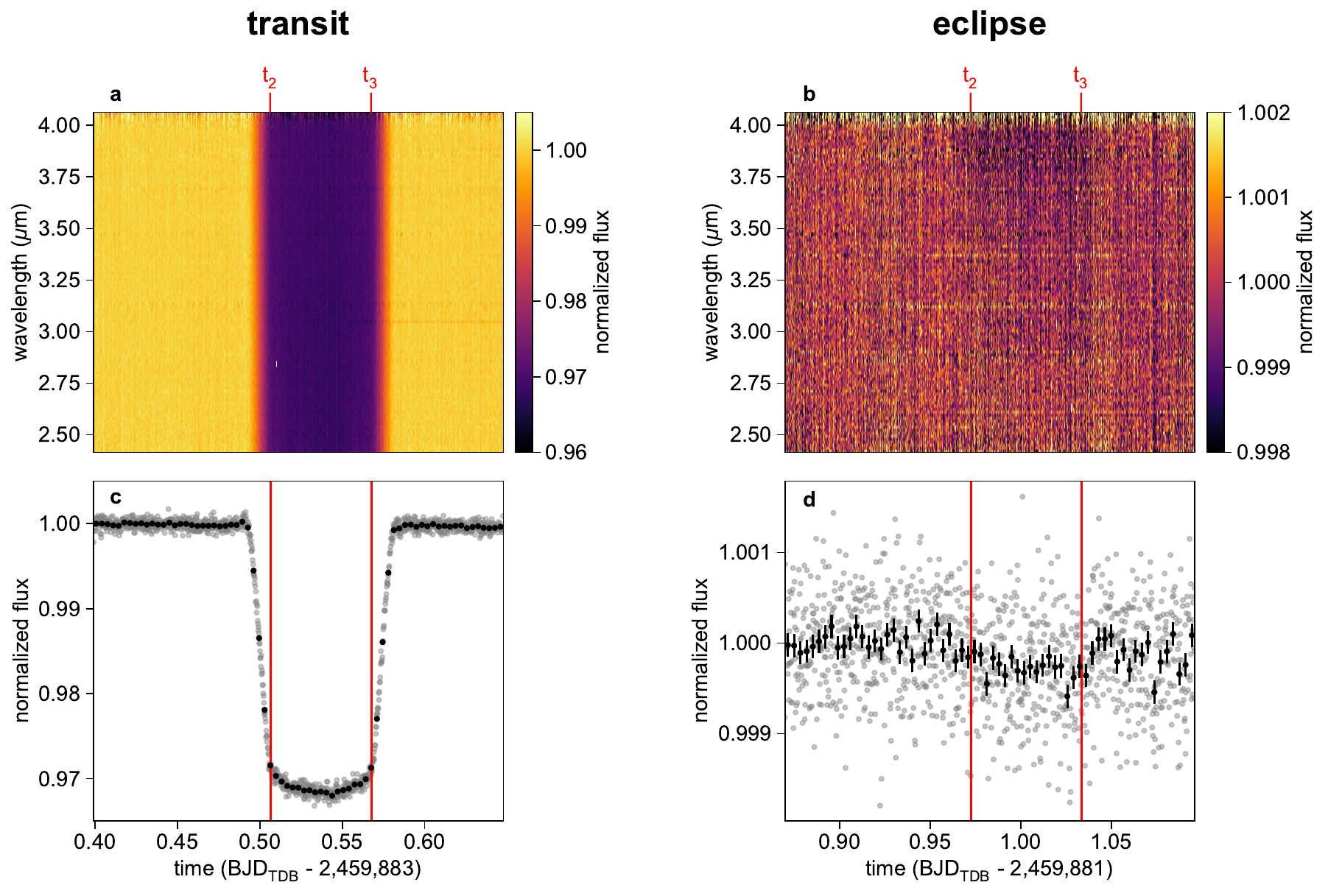}
    \caption{\textbf{Spectroscopic and broadband NIRCam F322W2 lightcurves of the transit and eclipse of WASP-80b.} The raw spectroscopic transit (\textbf{a}) and eclipse (\textbf{b}) lightcurves after spectral binning (0.0146\,$\mu$m bins) but without any temporal binning. The transit and eclipse's second and third contact points are indicated with red lines above the plots. The broadband (2.420--4.025\,$\mu$m) transit (\textbf{c}) and eclipse (\textbf{d}) lightcurves are shown in gray without error bars. The transit and eclipse's second and third contact points are indicated with red vertical lines. Black points with 1$\sigma$ error bars show temporally binned data with a cadence of 5 minutes; note that the error bars in panel c are smaller than the point size. BJD$_{\rm TDB}$ is the date in the Barycentric Julian Date in the Barycentric Dynamical Time system.}
    \label{fig:lightcurvesMain}
\end{figure*}

We observed the eclipse and transit of WASP-80b using the F322W2 grism of JWST's Near Infrared Camera (NIRCam)\cite{NIRCam:2004} on the 28th and 29th of October, 2022 UTC, respectively. Each observation consisted of 1227 integrations with 6 groups per integration and used the BRIGHT2 readout pattern and the SUBGRISM256 subarray, giving an overall exposure duration of 5.97 hours per observation. We analyzed the spectra from both observations using two independent pipelines (\texttt{Eureka!}\cite{bell2022} and \texttt{tshirt}\cite{tshirt:2022}) to ensure our results were robust to different data reduction and fitting methods. The raw broadband (2.420--4.025\,$\mu$m) and spectroscopic lightcurves are shown in Fig.~\ref{fig:lightcurvesMain}.

\begin{figure}
    \centering
    \includegraphics[width=\linewidth]{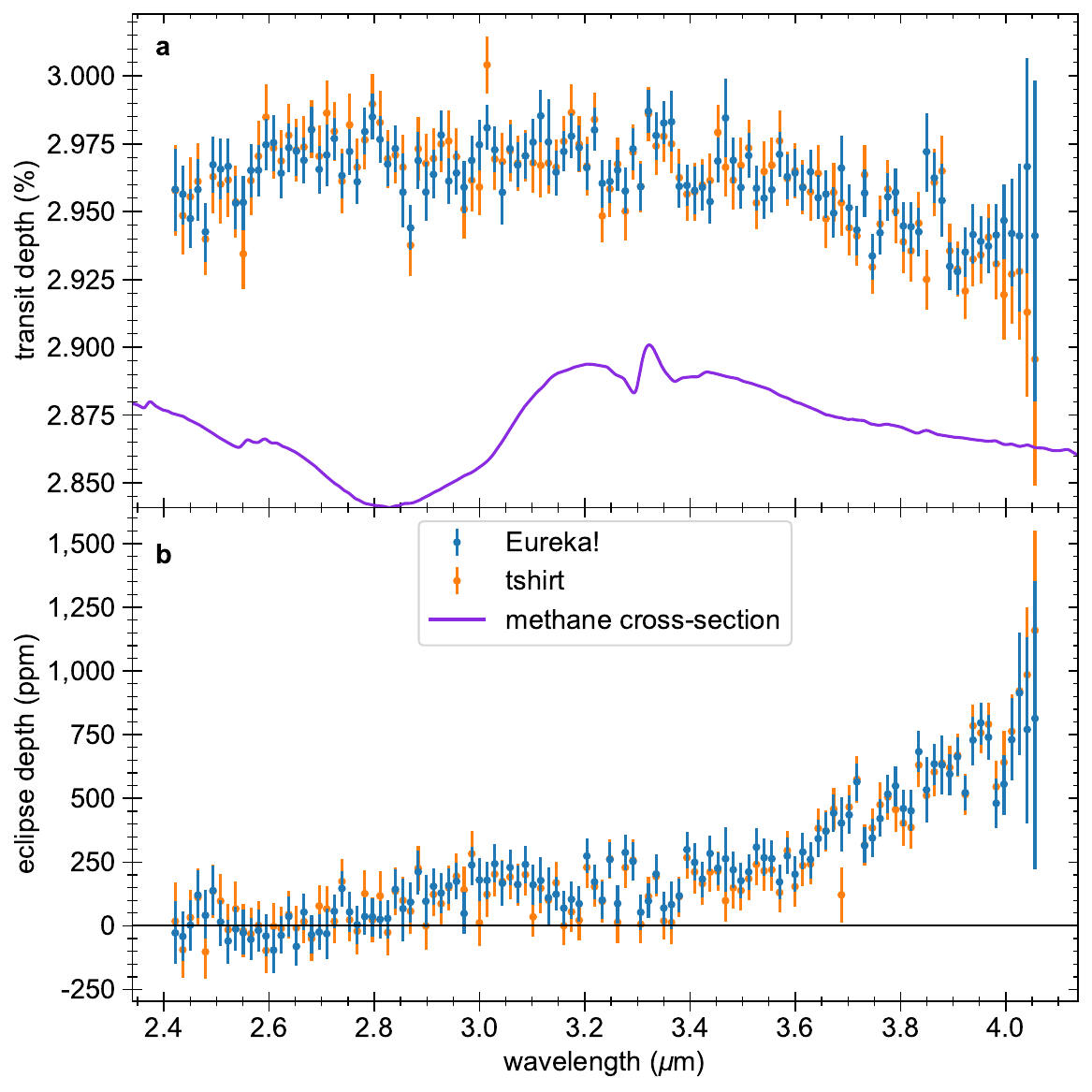}
    \caption{\textbf{Independent reductions of the WASP-80b transmission and emission spectra.} \textbf{a}, The transmission spectrum produced by \texttt{Eureka!} compared to that produced by \texttt{tshirt} after offsetting the \texttt{tshirt} spectrum up by a constant 161\,ppm. After the constant offset, the two spectra agree well within uncertainties (1$\sigma$ error bars are shown). Also shown in red is the log of the cross-section of methane (in arbitrary units) to help indicate where methane signatures are expected. \textbf{b}, The emission spectrum produced by \texttt{Eureka!} and \texttt{tshirt} which agree well within uncertainties. Neither analysis enforced a prior that the eclipse depths be greater than zero, but a solid line at 0\,ppm shows that all wavelengths are consistent with a 0\,ppm or greater eclipse depth at 1$\sigma$ (1$\sigma$ error bars are shown).}
    \label{fig:spectraComparisons}
\end{figure}

We find that our independent analyses agreed well within uncertainties (Fig.~\ref{fig:spectraComparisons}) for the emission spectra, with both showing suppressed emission longward of $\sim$3.1\,$\mu$m indicating the presence of methane. There was a constant offset between the two transmission spectra with the \texttt{Eureka!} median transit depth being 161\,ppm deeper than the median transit depth from \texttt{tshirt}. We ultimately choose to offset the \texttt{tshirt} spectra by +161\,ppm to match the \texttt{Eureka!} spectra (discussed in detail in the Methods), after which the two spectra agree well within uncertainties (Fig.~\ref{fig:spectraComparisons}). Both transmission spectra also show clear evidence for methane with a `W'-shaped absorption feature centered around 3.3\,$\mu$m composed of the central ro-vibrational Q branch and the adjacent P and R branches. For both the transmission and emission spectra, we then use the mean of the two independent analyses as our fiducial spectra and inflate our uncertainties to account for the disagreement between the two analyses (see Methods).

\begin{figure}
    \centering
    \includegraphics[width=\linewidth]{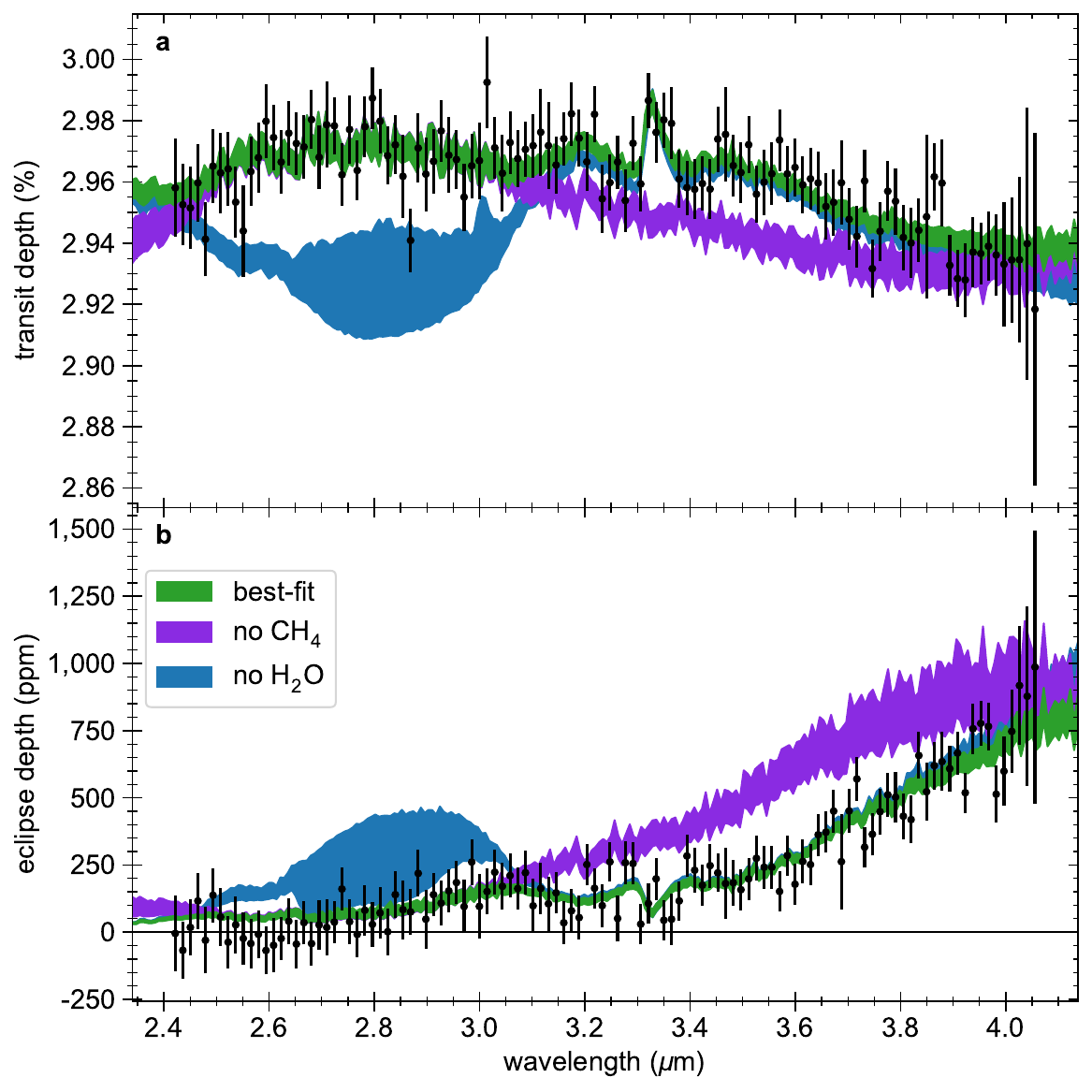}
    \caption{\textbf{Interpretation of WASP-80b's transmission and emission spectra.} The observed transmission (\textbf{a}) and emission (\textbf{b}) spectrum  (with 1$\sigma$ error bars) are compared to the best-fit free retrieval model. Colored regions demonstrate the contribution of different molecules, demonstrated with 68\% confidence intervals for the ``best-Fit" free retrieval (green), with ``no CH$_4$'' opacity (purple), and with ``no H$_2$O'' opacity (blue). Both the transmission and emission spectra are strongly shaped by CH$_4$ and H$_2$O.}
    \label{fig:retrievalMain}
\end{figure}

To interpret the observations and to quantify the detection of methane, we fit the emission and transmission spectra using a series of commonly employed atmospheric Bayesian inference methodologies known as retrievals\cite{Madhusudhan2018}. First, we perform retrievals using Aurora\cite{welbanks2021} on both the emission and transmission observations. Aurora computes atmospheric models assuming absorption due to H$_2$O, CH$_4$, NH$_3$, CO, CO$_2$, and SO$_2$ with constant-with-altitude gas volume mixing ratios, a non-isothermal parameterized pressure-temperature profile following the prescription from ref.\cite{MS09}, H$_2$-H$_2$ and H$_2$-He collision-induced absorption, and inhomogeneous clouds and hazes (for transmission only, see Methods). The Bayesian parameter estimation is performed using the nested sampling algorithm MultiNest\cite{Feroz2013}.

Our retrievals on the transmission and emission spectra find that the atmosphere of WASP-80b is best explained by significant absorption due to CH$_4$ and H$_2$O, while NH$_3$, CO, CO$_2$, and SO$_2$ are not constrained by these observations. Both emission and transmission spectra show a significant absorption feature of CH$_4$ between 3.0$\mu$m and 3.6$\mu$m, with its distinctive 3.3$\mu$m absorption line being clearly visible in both spectra. The spectra and their associated fits are summarized in Fig.~\ref{fig:retrievalMain}. The retrieved chemical abundances are consistent between our emission (i.e., $\log_{10}(\rm{CH}_4){=}{-}3.9^{+0.6}_{-0.7}$) and transmission data (i.e., $\log_{10}(\rm{CH}_4){=}{-}4.2^{+0.5}_{-0.7}$). The consistency between our dayside and terminator methane abundances is in agreement with theoretical predictions for warm exoplanets which are expected to show minimal longitudinal temperature gradients and well-mixed atmospheres\cite{perez-becker2013,cooper2006}. We investigate the significance of this molecular absorption feature by performing a Bayesian nested model comparison. Our model comparison results in a significant 6.1$\sigma$ detection of CH$_4$ in both the transmission and emission spectra (for $\sigma$ estimation, see Methods).

This strong and significant detection of CH$_4$ is confirmed by our second modeling approach. Here, we compute a grid of self-consistent 1-dimensional radiative-convective photo-chemical-equilibrium (1D-RCPE) models and fit for the atmospheric metallicity, carbon-to-oxygen ratio, irradiation temperature, and the presence of an opaque gray cloud opacity. While this method rules out implausible chemical/thermal solutions due to the inherent 1D-RCPE assumption, it is also less flexible due to the fewer number of free parameters in the paradigm. Within this framework, CH$_4$ is detected at a higher significance of 8.1$\sigma$ in transmission and 8.7$\sigma$ in emission. Our 1D-RCPE fit to the transmission spectrum suggests metallicities between $\sim$3--10$\times$ solar (at $\pm$1$\sigma$) and a roughly solar to sub-solar C/O ratio (C/O $< 0.57$ at 2$\sigma$), while the fit to the emission spectrum is largely uninformative on composition. This is generally consistent with the expectation of ${\sim}5\times$ solar metallicity for the planet from proposed relationships between mass and atmospheric metallicty\cite{Kreidberg+14b,Welbanks+19b}. Further, ref.\cite{Thorngren+16} determined a bulk metal mass fraction of $Z_{\rm p}=0.12$, which, when combined with the proposed relationship between bulk metal mass fraction and atmospheric metallicity of ref.\cite{Bean+23}, would suggest a ${\sim}4\times$ solar atmospheric metallicity which is also consistent with our findings.

The dearth of robust methane detections to date has prompted theoretical studies into mechanisms to deplete methane in exoplanet atmospheres, such as high metallicity, low C/O, and high interior heat flux\cite{madhusudhan2011b,moses2013,morley2017,benneke2019,fortney2020}. However, this definitive detection of methane throughout the atmosphere of WASP-80b with low-resolution JWST spectroscopy raises the question to what extent past non-detections were affected by the sparse wavelength coverage and precision achievable with \textit{HST} and \textit{Spitzer}. Future studies from the MANATEE program will explore the transmission and emission spectrum of WASP-80b from 4--12\,$\mu$m which will provide more precise compositional constraints for the planet, and further MANATEE program observations will target other warm exoplanets, including those where \textit{HST} and \textit{Spitzer} were unable to detect methane (GJ 436b and GJ 3470b) and others we may expect to contain methane (WASP-69b, WASP-107b). Methane has been the gas species used to constrain the carbon abundance of solar system giant planets\cite{PPIV2022}, and with this robust \textit{JWST} detection we can begin to directly link warm exoplanets and solar system giant planets under a unifying measurement. Precise measurement of the methane mixing ratios in WASP-80b and other planets will allow for more quantitative analyses of their atmospheric compositions and formation pathways\cite{oberg2011co}, and will also permit more thorough investigations into disequilibrium processes.



\clearpage
\section*{Methods} \label{sec:methods}
\renewcommand{\figurename}{Extended Data Fig.}
\renewcommand{\tablename}{Extended Data Table}
\renewcommand{\thefigure}{Extended Data Fig.~\arabic{figure}}
\renewcommand{\theHfigure}{Extended Data Fig.~\arabic{figure}}
\renewcommand{\thetable}{Extended Data Table \arabic{table}}
\renewcommand{\theHtable}{Extended Data Table \arabic{table}}
\setcounter{figure}{0}
\setcounter{table}{0}

\subsection*{Data Reduction} \label{sec:reduction}

\paragraph{Eureka!}\label{sec:eurekaReduction}

Our \texttt{Eureka!} reduction used version 0.9 of the \texttt{Eureka!} pipeline\cite{bell2022}, CRDS version 11.16.16 and context 1019, and \texttt{jwst} package version 1.8.3\cite{jwst_v1.8.2}. The \texttt{Eureka!} Control Files and \texttt{Eureka!} Parameter Files we used are available for download (\url{https://doi.org/10.5281/zenodo.8264964}), and the important parameters are summarized below.

\texttt{Eureka!}'s Stage 1 and 2 use the \texttt{jwst} pipeline for basic calibration, and both stages were run with their default settings with the exception of increasing the Stage 1 jump step's rejection threshold to 6.0 (to avoid excessive false-positives) and skipping the photom step in Stage 2 as we do not need photometrically calibrated data. Our Stage 3 settings largely follow those used for the \texttt{Eureka!} spectroscopic reduction of ref.\cite{ahrer2022nircam}, with some slight modifications. In particular, we instead cropped each frame to only include y-pixels 5--64 and x-pixels 15--1709, masked pixels marked as ``DO\textunderscore NOT\textunderscore USE'' in the data quality (DQ) array, clipped background pixels at 5-sigma along the time axis, and smoothed the median frame used for optimal spectral extraction\cite{horne1986optspec} along the spectral direction using a 13-pixel wide boxcar filter, and we compute the temporal evolution of the spatial position and PSF-width (after first summing along the spectral direction) to later decorrelate against. We then binned the spectra into 113 spectral channels spanning 2.420--4.025\,$\mu$m (0.0146\,$\mu$m bins), and we also produced a single ``broadband" lightcurve spanning 2.420--4.025\,$\mu$m. We then removed a small number of obvious $>$10-sigma outliers compared to high-pass filtered version of the data computed using a 50-integration wide boxcar filter (to ensure we do not sigma-clip the transit itself). We also removed the first 110 integrations from the eclipse visit which showed a decreasing signal.

\paragraph{tshirt}\label{sec:tshirtReduction}

The \texttt{tshirt} pipeline is an open source pipeline that extracts lightcurves of spectroscopic and photometric data (\url{https://tshirt.readthedocs.io}).
In the \texttt{tshirt} analysis, we modified the CALDETECTOR1 stage of the \texttt{jwst} pipeline to turn the \texttt{\_uncal} files into count rate images with less 1/$f$ noise.
We started with \texttt{jwst} pipeline version 1.6.0\cite{jwst_v1.8.2}, CRDS Version 11.16.5, and CRDS context \texttt{jwst\_1009.pmap}.
Instead of the default reference pixel correction, we use a row-by-row, odd/even by amplifier (ROEBA) correction\cite{schlawin2023} using the background pixels, which can reduce the 1/$f$ noise as compared to reference pixels alone\cite{schlawin2020jwstNoiseFloorI}.
We use the reference pixels in the bottom 4 rows for the odd/even correction and all pixels from X=1846 to 2043 (0-based) for the row-by-row correction.
For the jump step of the pipeline, we use a threshold of 6\,$\sigma$.
We then proceed with the rest of the steps in CALDETECTOR1 with the default parameters.
We manually divide the images by an \texttt{jwst\_nircam\_flat\_0266.fits} imaging flat field and mark all pixels that have a ``DO\textunderscore NOT\textunderscore USE'' DQ value to NaN.
We multiply all rate images by the gain and integration time to estimate the total number of electrons at the end of the ramp.
Finally, we clean all images by combining 150 rate images at a time and mark all pixels that deviate from the median rate image by more than 20 times the jwst pipeline error for a given rate image as NaN.

We perform a column-by-column background subtraction with a linear fit along the Y direction to pixels Y=5 to 24 and Y=44 to 65 for all rate images.
For the spectral extraction, we first fit the spectrum row-by-row along the dispersion direction of integration 613 with a smooth 40 knot univariate spline with SciPy\cite{scipy} in order to build a profile function as a function of wavelength.
We then use a co-variance weighted extraction\cite{schlawin2020jwstNoiseFloorI}, assuming a uniform correlation between pixels of 0.08 and a read noise of 14\,$e^-$.
We extract a rectangular aperture from 29 to 39 pixels in the Y direction and 4 to 1747 in the X direction.
While this rectangular aperture does not track the curvature of the trace and could influence the absolute flux and aperture corrections, we normalize the spectra by the out-of-transit and out-of-eclipse baselines, so this effect is not expected to affect the derived planet spectra.
We use the profile along the Y direction to estimate the missing flux in pixels that marked as NaN and which deviate from the average profile by $\geq$120$\sigma$.

\subsection*{Lightcurve Fitting} \label{sec:fitting}

\paragraph{Eureka!}\label{sec:eurekaFitting}

We first fit the transit broadband lightcurve to get the best possible constraints on the planet's orbital parameters. We adopted Gaussian priors on the orbital period ($P$), linear ephemeris ($t_{\rm 0}$), inclination ($i$), and scaled semi-major axis ($a/R_{\rm *}$) based on the values of ref.\cite{triaud2015} (see also Extended Data Table \ref{tab:orbitalParameters}) and assumed zero eccentricity. We used a \texttt{starry}\cite{starry} transit model with broad, uninformative priors on the transit depth and freely fitted the reparameterized quadratic limb darkening parameters\cite{kipping2013}. We allowed for a linear trend in time, and linearly decorrelated against the changes in the spatial position and spatial PSF-width computed during \texttt{Eureka!}'s Stage 3. Finally, we also fitted a white-noise multiplier to ensure a reduced chi-squared of 1 and avoid artificially constrained posteriors. No mirror ``tilt'' events\cite{schlawin2023} were evident in either of our transit or eclipse observations.

We then sampled the posteriors of our model using \texttt{PyMC3}'s No U-Turns Sampler\cite{pymc3} using two independent chains, each taking 7,000 tuning draws and then 3,000 posterior samples with a target acceptance rate of 0.95. We confirmed the chains had converged using the Gelman-Rubin statistic\cite{GelmanRubin1992} and a visual comparison of the two independent chains. We then used the 16th, 50th, and 84th percentiles from the \texttt{PyMC3} samples to estimate the best-fit values and their uncertainties. The orbital parameters derived from this fit are tabulated in Extended Data Table \ref{tab:orbitalParameters}. We find a white-noise level 6.1 times larger than the expected photon limit for this fit to the broadband lightcurve; the excess noise is primarily caused by 1/$f$ noise which degrades the improvements seen by spectrally binning the data\cite{schlawin2020jwstNoiseFloorI}. There is no evidence for stellar activity in either observation nor any starspot crossings in the transit observations which is consistent with past transit observations\cite{fukui2014,mancini2014}, although the star has previously been estimated to have a $\sim$3\% spot-covering fraction assuming a spot temperature $\sim$500\,K below the stellar effective temperature\cite{kirk2018,wong2022}.

For the transmission spectrum, we then froze the orbital parameters to the median values from the transit broadband lightcurve. We otherwise fit the data in the same manner as the broadband lightcurve, with the exception of decreasing the number of tuning steps to 3000, the number of posterior samples to 1000, and the target acceptance rate to 0.85 as the sampler had fewer difficulties converging for these less precise observations. For the emission spectrum, we also froze the orbital parameters as well as the planetary radius to the median values from the transit broadband lightcurve fit. We otherwise fit the data in the same manner as the transmission spectrum with the exception of using \texttt{starry}'s eclipse model (assuming a uniform brightness map for the planet). For these spectroscopic fits, we find the noise level to typically be $\sim$25\% above the photon limit; the main exception is in the last $\sim$5 wavelength bins where the noise climbs to 2.2 times the photon limit as this is where the throughput of the F322W2 filter drops off steeply and we become more strongly sensitive to pointing jitter in the spectral direction. As shown in Extended Data Fig.~\ref{fig:AllanPlots}, our spectroscopic fits show no evidence for residual red noise.

\paragraph{tshirt}\label{sec:tshirtFitting}

Our \texttt{tshirt} analysis followed nearly the same procedure as the \texttt{Eureka!} analysis and also used the \texttt{starry} lightcurve code for the emission spectrum and \texttt{exoplanet} code for the transmission spectrum.
As with the \texttt{Eureka!} analysis, we started with Gaussian priors on the orbital parameters from ref.\cite{triaud2015} on the broadband lightcurve fits and uninformative quadratic limb darkening parameters\cite{kipping2013}.
We first bin both the broadband and spectroscopic lightcurves into 300 time bins for faster lightcurve evaluation.
We find the maximum a priori solution to the lightcurves and clip any points that are more than 5$\sigma$ from the maximum a priori lightcurve and then find a new maximum a priori solution for the clipped lightcurve points.
We fit the lightcurves with a second order baseline (i.e., a normalization term, a linear term, and quadratic term with time, where the time coordinate is normalized to be between -1.0 at the start and 1.0 at the end of the time series).
We also use the \texttt{pymc3} No U-Turns sampler.
We initialize the fit from the maximum a priori solution and sample 3000 tuning steps and 3000 posterior draws for the broadband lightcurve and spectroscopic fit.
As with \texttt{Eureka!} we froze the orbital parameters from the broadband (at the mean of the posterior distribution; see Extended Data Table \ref{tab:orbitalParameters}) and use this for all spectroscopic parameters.
The final transmission and emission depths and errors are reported as the mean and standard deviation of the posterior distribution, respectively.

We compared the measured standard deviation of the lightcurve from integrations 25--401 (the out-of-transit baseline) to the theoretical photon and read noise.
We found that the broadband standard deviation per integration was 365\,ppm as compared to 93\,ppm theoretical limit, most likely due to 1/$f$ noise.
The individual spectroscopic lightcurves ranged from 0.99 to 1.15 times the theoretical limit (5th to 95th percentile) and exhibited minimal residual red noise (see Extended Data Fig.~\ref{fig:AllanPlots}).

\paragraph{Comparing and Combining Independent Spectra}\label{sec:combiningSpectra}

As shown in Fig.~\ref{fig:spectraComparisons}, aside from a constant 161\,ppm offset between the \texttt{Eureka!} and \texttt{tshirt} transmission spectra, our two analyses were in close agreement at nearly all wavelengths in both the transmission and emission spectra, very rarely disagreeing by more than 1$\sigma$. In the hopes of eliminating the constant offset between our transmission spectra, we confirmed that the limb darkening parameterization was the same between both analyses and confirmed that both our priors on the planet-to-star radius ratio were minimally informative. Next, we confirmed that the orbital parameters derived by each pipeline were consistent with each other and also investigated the impact of different systematic models during the fitting stage between the two analyses. We refitted the \texttt{Eureka!} reduction using the exact same orbital parameters and systematic model as used in the \texttt{tshirt} analysis and found that the resulting transmission spectrum was still offset from the \texttt{tshirt} transmission spectrum by 153\,ppm. We also investigated the use of group-level background subtraction (GLBS) which the \texttt{tshirt} analysis used but the \texttt{Eureka!} analysis did not use. We re-reduced the data with \texttt{Eureka!} using GLBS and using the \texttt{tshirt} orbital parameters and systematics model when fitting, but the resulting spectrum was still offset from the \texttt{tshirt} analysis by 145\,ppm. Finally, we performed a test with the ROEBA step turned off in the \texttt{tshirt} pipeline; the resulting spectra were 157\,ppm deeper than the fiducial \texttt{tshirt} transmission spectra, putting them right in line with the \texttt{Eureka!} transmission spectra. While the row-by-row subtraction algorithm\cite{schlawin2020jwstNoiseFloorI,schlawin2023} used by \texttt{tshirt} reduces 1/$f$ noise, it appears that it also either introduces or removes a bias in the mean transit depth. As the mean transit depth was ultimately not of importance for this work, we chose to offset the \texttt{tshirt} transmission upward by 161\,ppm so that the median transit depth between both pipelines were in agreement, and we leave a detailed investigation into the possible biases introduced or removed by the row-by-row subtraction algorithm for future work.

For our fiducial spectra, we decided to average together our two analyses (after offsetting \texttt{tshirt}'s transmission spectrum) and inflate our uncertainties to reflect any disagreements between the two. When combining our independent spectra, we used the mid-point between \texttt{Eureka!} and \texttt{tshirt} at each wavelength as our mean depth. For our final uncertainties, we used the mean uncertainty between \texttt{Eureka!} and \texttt{tshirt} and then added in quadrature an error inflation term equal to the root-mean-squared difference between the analyses at each wavelength. In the case of only two pipelines, this error inflation term simplifies to just half the difference between the \texttt{Eureka!} and \texttt{tshirt} values. This inflates our median uncertainty by only 3\% but appreciably inflates the uncertainties at the few wavelengths where larger disagreements arise between \texttt{Eureka!} and \texttt{tshirt}. Finally, we also performed retrievals on the individual \texttt{Eureka!} and \texttt{tshirt} spectra and found that the results were consistent with the retrievals on our fiducial, combined spectra.

\subsection*{Atmospheric Retrievals} \label{sec:retrievals}
To interpret the emission and transmission spectra, we use atmospheric radiative transfer models combined with a Bayesian inference method--commonly referred to as ``atmospheric retrievals".  Retrievals employ a parameterized model of the atmosphere (a ``forward model"), typically a 1D column, that includes molecular gas abundances, a pressure-temperature profile, and some flavor of cloud opacity in order to compute an emission or transmission spectrum.  Below we describe two such forward models that span a range of assumptions. Parameter estimation with both of these models is performed using the nested sampling algorithm\cite{Skilling2006} Multinest\cite{Feroz2013} with the {\tt pymultinest} implementation\cite{Buchner2014}, which derives the posterior-probability distribution (constraints on each model parameter and their correlations) as well as the Bayesian evidence ($\mathcal{Z}$) for each model. The ratio in model evidences (the Bayes factor, $B_{12}=\mathcal{Z}_1/\mathcal{Z}_2$) can be used to compare two models and estimate the preference for one over another. This preference for a given model parameter can be converted to a corresponding $\sigma$ level and is what we use to determine the detection significance of methane\cite{trotta2008,benneke2013,welbanks2021}.

\paragraph{Free Retrieval}\label{sec:LuisRetrieval}
We begin by performing agnostic atmospheric model inferences with no assumptions about the chemical composition of the planet or its pressure-temperature profile. Instead of assuming chemical equilibrium or radiative-convective conditions, this modeling paradigm attempts to understand the state of the atmosphere by fitting directly for the atmospheric conditions through a series of parameters for its chemical composition and its pressure-temperature profile with no a priori expectations of physical consistency. This modeling approach is commonly known in the field as `free retrievals'\cite{Fortney2021}. We use the retrieval framework Aurora\cite{welbanks2021}, a code recently expanded from its original applications to transmission spectroscopy of exoplanets\cite{Welbanks2022} to the modeling and interpretation of emission spectra of transiting exoplanets. Aurora includes the general methods for modeling emission spectra of transiting exoplanets as described in refs.\cite{Line2013ApJ,Gandhi2018}. We only consider secondary eclipse models in the pure absorption limit with negligible scattering into the beam of radiation, an approximation generally appropriate for the wavelengths of these JWST observations. 

The equation of radiative transfer for the emission spectrum is calculated by modeling the planetary atmosphere using $N$ layers or slabs, where each slab $i$ has an optical depth $\tau$, a temperature $T$, and a radiation intensity $I_{i-1}$ from the layer underneath at an angle $\theta$ from the normal such that $\cos(\theta)=\mu$. Then, the radiation emergent from each slab is 

\begin{equation}
    I_i(\lambda, \mu) = I_{i-1}(\lambda, \mu) e^{-\tau(\lambda)/\mu} + B(T,\lambda)(1-e^{-\tau(\lambda)/\mu}),
\label{methods:RT_emission}
\end{equation}

\noindent where $B(T,\lambda)$ is a Planck function at the temperature $T$ and wavelength $\lambda$; ref.\cite{Seager2010}. The outgoing flux at the top of the atmosphere ($F_{\rm{top}}$) is calculated by integrating Equation \ref{methods:RT_emission}, which in this $N$ model layer we numerically integrate using the double ray method with the weights $W(\mu)$ and $\mu$ values from ref.\cite{Gandhi2018}. The observed planet-to-star flux ratio is given by the ratio of the planetary and stellar areas multiplied by the ratio of the outgoing flux at the top of the atmosphere and the stellar flux\cite{Seager2010}. We describe the stellar flux using an interpolated PHOENIX\cite{Huesser2013} stellar model at the published\cite{triaud2015} photospheric temperature, stellar metallicity, radius, and gravity. 

The description of the equations of radiative transfer for transmission spectroscopy as implemented in Aurora is available in ref.\cite{welbanks2021} and are solved assuming a parallel-plane atmosphere. As with the emission spectra models above, the transmission model assumes hydrostatic equilibrium, and constant-with-height atmospheric abundances. The resulting planet-to-star flux ratios (e.g., the resulting spectra) for the emission and transmission models are convolved and binned to the resolution of the NIRCam observations taking into account the instrument sensitivity and following the methods in refs.\cite{welbanks2021, Gandhi2018, Pinhas2018}. 

For both configurations, the atmospheric models assume line-by-line opacity sampling and are computed at a spectral resolution of R=30,000. The chemical absorbers considered are H$_2$O\cite{Rothman2010}, CH$_4$\cite{Yurchenko2014}, NH$_3$\cite{Yurchenko2011}, CO\cite{Rothman2010}, CO$_2$\cite{Rothman2010}, and SO$_2$\cite{Underwood2016}. The remaining filler gas is assumed to be a solar composition mixture of H$_2$+He, with a He to H$_2$ ratio of 0.17\cite{Asplund2009,welbanks2021}. Additionally, the models consider opacity due to H$_2$-H$_2$ and H$_2$-He collision induced absorption\cite{Richard2012}. The models also include the effects of H$_2$-Rayleigh scattering. For our transmission models with Aurora, we consider the possibility of inhomogeneous clouds and hazes using the 1 sector prescription of ref.\cite{welbanks2021}, in which the spectroscopic effects of clouds and hazes are combined with a clear atmosphere model following ref.\cite{line2016}. For the emission spectra, these considerations are not included in our fiducial model as the planet likely lacks clouds that scatter or emit at thermal infrared wavelengths\cite{fortney2005a, Madhusudhan2009}.

The pressure structure of the planet is discretized using 100 layers in both transmission and emission. The atmospheric extent in our emission models is from 100\,bar to 10$^{-6}$\,bar, while the transmission models extend from 100 bar up to 10$^{-9}$\,bar. The pressure-temperature structure is parameterized following the 6-parameter prescription from ref.\cite{Madhusudhan2009}, which has been widely adopted in the field to describe the thermal structure of exoplanetary atmospheres in both emission and transmission geometries. In our transmission retrievals, we infer the reference pressure (P$_{\rm{ref.}}$) at the reported planetary radius of WASP-80b. On the other hand, our emission models assume that this planetary radius corresponds to a pressure $\sim0.1$\,bar in order to calculate the equation of hydrostatic equilibrium. 

Our transmission model has a total of 17 parameters: 6 for the volume mixing ratios of the chemical species, 6 for the pressure-temperature profile, 4 for the inhomogeneous clouds and hazes, and 1 for the reference pressure. On the other hand, our emission models have 12 parameters: 6 for the volume mixing ratios of the chemical species and 6 for the pressure-temperature profile. We compute the Bayesian parameter estimation using pyMultiNest\cite{Buchner2014} as described above, and a series of nested retrievals (e.g., removing a single species from the full retrieval) to calculate the model preference for each chemical species in our model. 

The key results from the free retrievals are summarized in Extended Data Tables \ref{tab:atmoProperties} and \ref{tab:evidences} and Extended Data Fig.~\ref{fig:molecule_constraints}, and Extended Data Figs.~\ref{fig:aurora_transmission} and \ref{fig:aurora_emission} show corner plots showing the covariance between all fitted parameters. Our retrieved methane abundances are consistent between emission ($\log_{10}(\rm{CH}_4){=}{-}3.9^{+0.6}_{-0.7}$) and transmission data ($\log_{10}(\rm{CH}_4){=}{-}4.2^{+0.5}_{-0.7}$), with strong detections of CH$_4$ at 6.1\,$\sigma$ in both spectra. Besides CH$_4$, we find significant absorption due to H$_2$O in WASP-80b while the abundances of CO, CO$_2$, NH$_3$, and SO$_2$ are not constrained with either of these observations. The presence of H$_2$O is preferred at 2.6\,$\sigma$ with an abundance of $\log_{10}(\rm{H}_2\rm{O}){=}{-}2.2^{+0.8}_{-1.1}$ in emission and at 4.6\,$\sigma$ with an abundance of $\log_{10}(\rm{H}_2\rm{O}){=}{-}1.8^{+0.6}_{-1.0}$ in transmission. These abundances are consistent with each other, although aided by the $\sim$1\,dex constraints. Planets within this temperature regime are expected to have fairly homogeneous longitudinal temperature distributions ($\lesssim$200\,K at the typical pressures probed in transmission)\cite{kataria2016}. We would, therefore, expect little to no variation in the thermochemical methane abundance between day and night (and across the terminator). Furthermore, ref.\cite{cooper2006} showed that horizontal mixing homogenizes longitudinal abundances to those on the planetary daysides. The consistency between our dayside and terminator methane abundances simultaneously confirms these expectations as well as bolsters the confidence in our modeling analysis.

The retrieved pressure-temperature profile from the transmission spectrum is largely unconstrained and consistent in shape with isothermal profiles near the equilibrium temperature of the planet of $\sim$825\,K. The retrieved temperature at the 100\,mbar pressure level, relatively near the expected photosphere of the planet\cite{Welbanks2019a}, is T$_{100\,\rm{mbar}}{=}960^{+330}_{-140}$\,K from our transmission spectrum retrievals. The transmission spectra do not constrain the presence of hazes in WASP-80b, mainly due to the lack of data in the optical. As for the presence of clouds, the retrieval infers the presence of optically thick clouds at pressures of $\sim$0.2\,mbar or lower ($2\,\sigma$ upper limit), covering 15\% to 60\% of the planet terminator ($2\,\sigma$ interval).

On the other hand, the retrieval of the emission data infers tighter constraints on the pressure-temperature structure of the planet with no temperature inversions and with a temperature at 100\,mbar of T$_{100\,\rm{mbar}}{=}912^{+74}_{-96}$\,K. The inferred dayside temperature is consistent with the planet's 825\,K equilibrium temperature, in agreement with the expectation that warm exoplanets should efficiently transport heat from their tidally synchronized daysides to their nightsides.

\paragraph{1D-RCPE Retrieval}\label{sec:MikeRetrieval}
Rather than fitting directly for the molecular gas abundances and pressure-temperature profile, we instead predict them under the assumption of 1-dimensional radiative-convective-photochemical-equilibrium (1D-RCPE) given the irradiation and elemental composition. Radiative-convective equilibrium (RCE) is used to compute the vertical pressure-temperature profile given the incident stellar flux, internal heat flux, and opacity structure of the atmosphere. RCE is achieved when the net flux-divergence (in the vertical direction) is zero\cite{marley2015}.  We use the ScCHIMERA RCE solver first described in ref.\cite{pis18}; with more recent updates given in refs.\cite{Mansfield2021,IyerLine2023}. Photochemical-equilibrium (PE) describes the ``disequilibrium" chemical state of the atmosphere accounting for the chemical kinetics arising from photochemistry and vertical mixing. We use the VULCAN\cite{Tsai2017,Tsai2022} tool to solve the photochemical kinetics problem to derive the vertical gas volume mixing ratios. We also assume an intrinsic temperature of 100\,K, following theoretical predictions\cite{thorngren2019}.

The SCHIMERA RCE solver and the VULCAN chemical kinetics solver are coupled to self-consistently derive the RCE pressure-temperature profile and the disequilibrium vertical gas mixing ratios given the incident stellar flux (re-scaled to a specified irradiation temperature, $T_{\rm irr}$, that then scales to the top-of-atmosphere PHOENIX\cite{Huesser2013} stellar flux spectrum), the atmospheric metallicity, $[M/H]$ (where ``[\,]" indicates log$_{10}$(M/H) relative to solar; 0 indicates solar, +1 indicates 10$\times$ solar, and so on), and the carbon-to-oxygen ratio, C/O (where the solar value is 0.46; \cite{Lodders2009}). The metallicity term uniformly scales the solar\cite{Lodders2009} abundances of all elements heavier than H and He (accounting for the re-normalization of H). The C/O is then adjusted after this scaling such that the sum of C+O is preserved. The gas mixing ratios are initialized under the assumption of thermochemical equilibrium (using the NASA CEA2 Gibbs free energy minimization solver\cite{Gordon1994}) given the pressure-temperature profile and elemental abundances. While the equilibrium chemistry routine solves for thousands of molecular/atomic species, we include the opacity sources for only the major radiative species over most exoplanet conditions (H$_2$/He collision-induced absorption, H/e-/H- bound/free-free continuum, and the line opacities for H$_2$O, CO, CO$_2$, CH$_4$, NH$_3$, H$_2$S, PH$_3$, HCN, C$_2$H$_2$, OH, TiO, VO, SiO, FeH, CaH, MgH, CrH, ALH, Na, K, Fe, Mg, Ca, C, Si, Ti, O, Fe+, Mg+, Ti+, Ca+, C+).

A converged RCE-thermochemical-equilibrium (1D-RCTE) solution to the atmospheric structure is first obtained.  The resulting 1D-RCTE temperature and mixing ratio profiles for  H$_2$, He, H$_2$O, CO, CO$_2$, CH$_4$, NH$_3$, N$_2$, H$_2$S, HCN, and C$_2$H$_2$ are used to initialize the VULCAN kinetics solver (to set the elemental abundances). We use the latest VULCAN setup (H-C-O-N-S kinetics network) as described in ref.\cite{Tsai2022}. We use the MUSCLES\cite{MusclesI,MusclesII,MusclesIII} UV-stellar spectrum from GJ676A (an M0V, v23) as a proxy for WASP-80. A power law eddy diffusion profile is used as described in ref.\cite{Tsai2022} (scaled in accordance with ref.\cite{Komacek2019}, resulting in diffusivities between 5$\times 10^{6}$ and 7$\times 10^{9}$ cm$^2$/s). After convergence, we then extract the VULCAN-modified mixing ratio profiles for the above species and ``fix" them in the 1D-RCTE solver, where equilibrium chemistry is again assumed for all other species (e.g., Na, K, etc.). This ScCHIMERA-to-VULCAN-to-ScCHIMERA represents a single ``iteration". We do this once more and find that the pressure-temperature profile and gas mixing ratios do not change. This processes is repeated over a grid of $T_{\rm irr}$ (725--900\,K in steps of 15\,K), [M/H] (-0.5--2.0 in steps of 0.125), and C/O (0.05--0.75 in steps of 0.05) resulting in over 4,000 converged 1D-RCPE atmospheric structures.

We perform parameter estimation over this grid ($T_{\rm irr}$, [M/H], C/O) by post-processing the 1D-RCPE atmospheric structures through either a transmission or emission spectrum geometry routine at an R=100,000, and then top-hat binned to the data wavelength grid (including only the relevant opacities\cite{Tennyson2020, Rothman2010, Grimm2015}, H$_2$-H$_2$/He CIA\cite{Karman2019}, H$_2$O\cite{Polyansky2018}, CO\cite{Gordon2015}, CO$_2$\cite{huang2012}, CH$_4$\cite{Hargreaves2020}, NH$_3$\cite{Coles2019}, HCN\cite{Harris2006}, C$_2$H$_2$\cite{Chubb2020}, and H$_2$S\cite{Azzam2016}) within the nested sampler. The temperature and gas mixing ratio profiles are tri-linearly interpolated (using the scipy\cite{scipy} {\tt RegularGridInterpolator}) to any given set of $T_{\rm irr}$, [M/H], and C/O before they are then passed into the spectral calculation routines. Interpolating the atmospheric structure is more accurate than using a nearest neighbor or interpolating pre-computed spectra directly. Within these on-the-fly spectral calculations we add in a vertically uniform grey cloud opacity ($\kappa_{\rm cld,em/tr}$) (independent opacities in emission and transmission). This results in a total of 4 parameters for each geometry. However, for the transmission geometry we include two additional parameters---the cloud patchiness fraction along the terminator\cite{line2016} and a scaling to the planetary radius (referenced to 1\,bar pressure)---for a total of 6 parameters to fit the transmission spectrum. Within this framework, we then fit the emission and transmission spectra separately. Extended Data Fig.~\ref{fig:retrieval_grid} summarizes these fits, and corner plots showing the covariance between the fitted parameters are available in Extended Data Figs.~\ref{fig:1DRCPE_transmission} and \ref{fig:1DRCPE_emission}. These inferences are then re-run after removing the methane opacity (note, methane is not a free parameter here, so there is no change in the number of parameters) and the resultant evidences are used to compute the 1D-RCPE methane {\it opacity} detection significance. For the transmission and emission scenarios, we obtain, respectively, ln$B_{12}$=30.9 and 36.0, corresponding to an 8.1 and 8.7\,$\sigma$ detection of methane. H$_2$O is also detected at 8.2\,$\sigma$ in transmission and at 3.4\,$\sigma$ in emission (see Extended Data Table \ref{tab:evidences}). While methane is readily detected in both emission and transmission within the 1D-RCPE framework, the constraints on [M/H] and C/O are not particularly informative due to the relatively narrow wavelength range which leads to a degeneracy between C/O and [M/H] (see Extended Data Figs.~\ref{fig:1DRCPE_transmission} and \ref{fig:1DRCPE_emission}). Analysis of additional, broader-wavelength datasets could help to break this degeneracy and enable more precise constraints on the elemental composition.


\subsubsection*{Data Availability}
The data used in this paper are associated with JWST GTO program 1185 (PI Greene; observations 2 and 4) and will be publicly available from the Mikulski Archive for Space Telescopes (\url{https://mast.stsci.edu}) at the end of their one-year exclusive access period.\\

\subsubsection*{Code Availability}
We used the following codes to process, extract, reduce and analyse the data: STScI's JWST Calibration pipeline\cite{jwst_v1.8.2}, \texttt{Eureka!}\cite{bell2022}, \texttt{tshirt}\cite{tshirt:2022}, starry\cite{starry}, PyMC3\cite{pymc3}, and the standard Python libraries numpy\cite{numpy}, astropy\cite{astropy2013, astropy2018}, and matplotlib\cite{matplotlib}.\\


\vskip 1 cm
\backmatter

\noindent\textbf{Acknowledgements}
T.J.B.~and T.P.G.~acknowledge funding support from the NASA Next Generation Space Telescope Flight Investigations program (now JWST) via WBS 411672.07.05.05.03.02. L.W.~acknowledges support for this work provided by NASA through the NASA Hubble Fellowship grant \#HST-HF2-51496.001-A awarded by the Space Telescope Science Institute, which is operated by the Association of Universities for Research in Astronomy, Inc., for NASA, under contract NAS5-26555. M.R.L.~acknowledge NASA XRP award 80NSSC19K0446 and STScI grant HST-AR-16139. M.R.L.~and L.W.~acknowledge Research Computing at Arizona State University for providing HPC and storage resources that have significantly contributed to the research results reported within this manuscript. Funding for E.S.~was provided by NASA Goddard Spaceflight Center. NIRCam team members are supported by NAS5-02105, a contract to the University of Arizona. K.O.~was supported by JSPS Overseas Research Fellowship. We thank Karl Misselt and Matthew Murphy for feedback on an early draft of the manuscript.\\

\noindent\textbf{Author contributions}
T.J.B.~led the data analysis effort, contributed the Eureka! analyses, verified the observing parameters, and led the writing of the paper.
L.W.~led the modeling analysis effort, contributed to the analysis/interpretation of the spectra and contributed to the text.
E.S.~contributed to the modeling, observing specifications before JWST launch, and ``tshirt" data analysis.
M.R.L.~contributed to the text, conceptual direction of the manuscript, and modeling analysis/interpretation of the spectra.
T.P.G.~contributed to the scientific case for making the observations and led the observation planning; he also contributed to focusing the scientific content of the manuscript.
J.F.~helped to plan initial observations, contributed text to the draft, and provided comments.
K.O.~helped to interpret the results and contributed to the text of the paper.
V.P.~helped with the physical interpretation of the spectrum.
E.R.~provided comments on the manuscript.
L.S.W.~provided preliminary 1D grid models.
S.M.~used the PICASO atmospheric model to perform model fitting analysis on an early version of the spectra.
T.G.B.~contributed to the planning and execution of the observations, evaluation of the observational results, modeling the stellar SED, and editing the manuscript.
M.L.B.~played a lead role in designing and executing the commissioning and calibration of the NIRCam Instrument.
M.R.~lead the development and testing of NIRCam including the demonstration of time series observations during commissioning.
J.A.S.~led development of JWST observation planning capabilities for exoplanet transits for 2 years, as well as commissioning of the NIRCam instrument, and provided inputs on the manuscript.\\

\noindent \textbf{Competing interests} The authors declare no competing interests.\\

\noindent\textbf{Additional information}\newline
\textbf{Correspondence and requests for materials} should be addressed to \href{Taylor J. Bell}{mailto:bell@baeri.org}.\newline
\textbf{Reprints and permissions information} is available at \url{www.nature.com/reprints}.

\pagebreak
\begin{appendices}

\begin{table*}[!htbp]
    \centering
    \small
    \begin{tabular}{cccc}
        \hline & Prior & \texttt{Eureka!}'s Posterior & \texttt{tshirt}'s Posterior \\ \hline
        $P$ (days) & 3.06785234 $\pm$ 0.00000083 & 3.067851919 $\pm$ 0.000000031 & 3.067851945 $\pm$ 0.000000026 \\
        $t_0$ (BJD$_{\rm TDB}$) & 2,456,487.425006 $\pm$ 0.000025 & 2,456,487.425006 $\pm$ 0.000025 & 2,456,487.425006 $\pm$ 0.000025 \\
        $a/R_*$ & 12.63 $\pm$ 0.1 & 12.612 $\pm$ 0.036 & 12.643 $\pm$ 0.032 \\
        $i$ ($^{\circ}$) & 89.02 $\pm$ 0.1 & 88.890 $\pm$ 0.066 & 88.938 $\pm$ 0.059 \\ \hline
    \end{tabular}
    \caption{\textbf{WASP-80b's orbital parameters.} The orbital priors\cite{triaud2015} used by both pipelines when fitting the transit broadband lightcurve as well as the posteriors derived from both pipelines. BJD$_{\rm TDB}$ is the date in the Barycentric Julian Date in the Barycentric Dynamical Time system.}
    \label{tab:orbitalParameters}
\end{table*}

\begin{figure*}[!htbp]
    \centering
    \includegraphics[width=\linewidth]{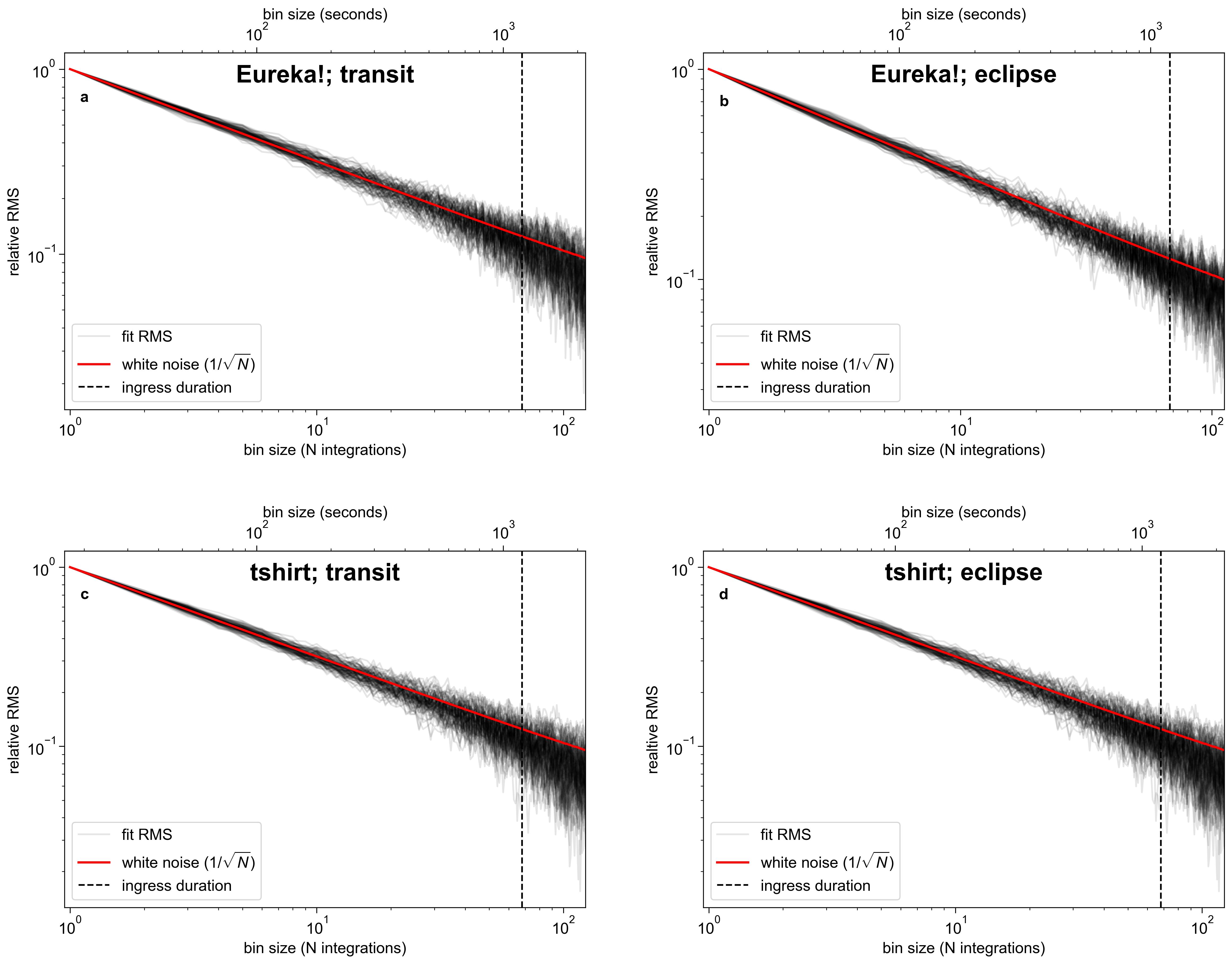}
    \caption{\textbf{Lack of residual red noise in spectroscopic fits.} Allan variance plots\cite{Allan1966} for each channel (normalized by the unbinned root mean square (RMS) in each channel) are shown in black lines while the ideal, white-noise behavior is shown in red. For reference, the timescale of transit/eclipse ingress is shown in each panel. The binned residuals closely follow the red line (within error) and indicate that there is no evidence for residual red noise in our fits.}
    \label{fig:AllanPlots}
\end{figure*}

\begin{table*}
    \centering
    \small
    \begin{tabular}{cl|cccc}
    \hline
     & Parameter  & Transmission & Emission & Prior Trans.  & Prior Emiss.  \\
    \hline
    \multirow{6}{*}{\rotatebox[origin=c]{90}{Chemical Species}}&$\log_{10} \left(X_{\textnormal{H}_2\textnormal{O}}\right)$&$-4.70^\Downarrow$ &$-8.55^\Downarrow$ &($-12,-1$) &($-12,-1$)  \\
     &$\log_{10}\left(X_{\textnormal{CH}_4}\right)$&$-4.2^{+0.5}_{-0.7}$ &$-3.9^{+0.6}_{-0.7} $&($-12,-1$) &($-12,-1$)   \\
    &$\log_{10}\left(X_{\textnormal{NH}_3}\right)$ &$-3.12^\Uparrow$&$-3.72^\Uparrow$&($-12,-1$) &($-12,-1$)  \\
    &$\log_{10}\left(X_{\textnormal{CO}}\right)$&$-1.32^\Uparrow$&$-1.05^\Uparrow$&($-12,-1$) &($-12,-1$)  \\
    &$\log_{10}\left(X_{\textnormal{CO}_2}\right)$&$-3.09^\Uparrow$&$-1.31^\Uparrow$&($-12,-1$) &($-12,-1$)  \\
    &$\log_{10}\left(X_{\textnormal{SO}_2}\right)$& $-4.68^\Uparrow$& $-6.84^\Uparrow$ &($-12,-1$) &($-12,-1$)  \\
    \hline
    \multirow{6}{*}{\rotatebox[origin=c]{90}{P-T}} &T$_{1\mu \text{bar}}$ (K) &$782 ^{+79}_{-138} $ &$545^{+56}_{-58}$ &$(0, 900)$ &$(300, 1200)$\\
    & $\alpha_1$~K$^{-1/2}$ &$0.9^{+0.6}_{-0.5}$ &$0.6^{+0.1}_{-0.1}$ &$(0.02, 2.00)$ &$(0.02, 2.00)$ \\
    &$\alpha_2$~K$^{-1/2}$  &$1.1^{+0.6}_{-0.5}$ &$0.9^{+0.7}_{-0.5}$ &$(0.02, 2.00)$ &$(0.02, 2.00)$  \\
    &$\log_{10}\text{P}_1$ (bar) &$-2.6^{+2.0}_{-2.3}$ &$ -0.1^{+0.9}_{-1.0}$ &$(-9, 2)$ &$(-6, 2)$  \\
    &$\log_{10}\text{P}_2$ (bar) &$-6.2^{+2.4}_{-1.8}$ &$-2.7^{+2.5}_{-2.1}$ &$(-2, 2)$ &$(-6, 2)$  \\
    &$\log_{10}\text{P}_3$ (bar) &$0.4^{+1.0}_{-1.3}$ &$1.2^{+0.6}_{-0.8}$ &$(-2, 2)$ &$(-6, 2)$  \\
    \hline
    &$\log_{10}\text{P}_\text{ref.}$  (bar)&$-3.5^{+0.9}_{-0.7}$ &N/A &$(-9, 2)$ &N/A   \\
    \hline
    \multirow{2}{*}{\rotatebox[origin=c]{90}{Cloud/Haze}} &$\log_{10}(a)$ &$2.6^{+4.2}_{-4.2}$ &N/A &$(-4, 10)$ &N/A  \\
    &$\gamma$ &$-10.0^{+6.6}_{-6.0}$ &N/A &$(-20, 2)$ &N/A \\
    &$\log_{10}(\text{P}_\text{cloud})$ (bar) &$-6.9^{+1.6}_{-1.3}$ &N/A &$(-9, 2)$ &N/A  \\
    &$\phi_\text{clouds and hazes}$ &$0.4^{+0.1}_{-0.1}$ &N/A &$(0, 1)$ &N/A\\
    \hline
    \end{tabular}
    \caption{\textbf{The retrieved atmospheric properties.} Retrieved parameters are either shown with their median retrieved values and 1$\sigma$ uncertainties or with just their 3$\sigma$ upper ($\Uparrow$) or 3$\sigma$ lower ($\Downarrow$) limits. The uniform prior range for each parameter are shown as (lower, upper) or as N/A when the parameter was not included in the retrieval.}
    \label{tab:atmoProperties}
\end{table*}

\begin{table*}
    \centering
    \small
    \begin{tabular}{c|cc|cc}
        \hline & \multicolumn{2}{c|}{Free Retrieval} & \multicolumn{2}{c}{1D-RCPE Grid-based Retrieval} \\
        & $\Delta\ln{\mathcal{Z}}$ & Significance ($\sigma$) & $\Delta\ln{\mathcal{Z}}$ & Significance ($\sigma$) \\ \hline
        CH$_4$ (transmission) & 16.6& 6.1 & 30.9 & 8.1 \\
        CH$_4$ (emission) & 16.8 & 6.1 & 36.0 & 8.7 \\ \hline
        H$_2$O (transmission) & 9.0 & 4.6 & 31.0 & 8.2 \\
        H$_2$O (emission) & 2.1 & 2.6 & 4.3 & 3.4 \\ \hline
    \end{tabular}
    \caption{\textbf{The significance of molecular detections.} The difference in the log Bayesian evidence ($\ln{\mathcal{Z}}$) between models with and without a specific molecule's opacity are shown for both the `Free' and `1D-RCPE' retrievals, and the corresponding detection significances are also shown.}
    \label{tab:evidences}
\end{table*}

\begin{figure*}
    \centering
    \includegraphics[width=\linewidth]{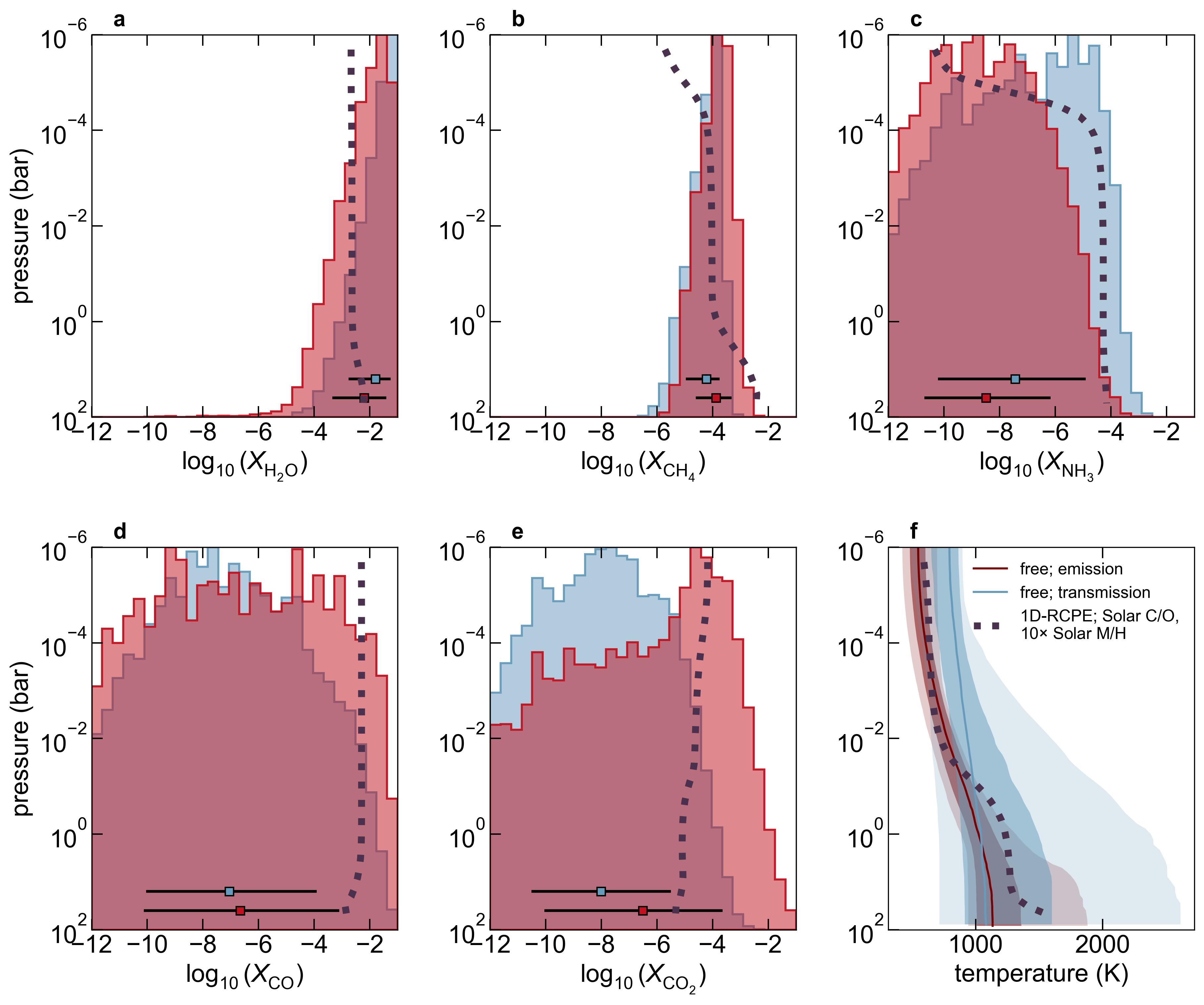}
    \caption{\textbf{Constraints on the abundances of key molecules in the atmosphere of WASP-80b.} In panels \textbf{a}--\textbf{e}, the retrieved abundances of some key molecules from the emission and transmission spectra are shown with red and blue histograms, respectively; these histograms are also summarised using similarly coloured points at the median of the histogram with 1\,$\sigma$ error bars (and placed at arbitrary pressure levels). Besides CH$_4$ whose abundance is bounded, all other posteriors are either upper or lower limits.
    In each of panels \textbf{a}--\textbf{e}, a representative 1D-RCPE derived mixing ratio profile with Solar C/O and 10$\times$ Solar metallicity\cite{Asplund21} is also plotted with a black dotted line. The retrieved pressure-temperature profile for each observing geometry is also shown in panel \textbf{f}, and the 1D-RCPE pressure-temperature profile corresponding to the model lines in panels \textbf{a}--\textbf{e} is also shown with a black dotted line.}
    \label{fig:molecule_constraints}
\end{figure*}

\begin{figure*}[!htbp]
    \centering
    \includegraphics[width=\linewidth]{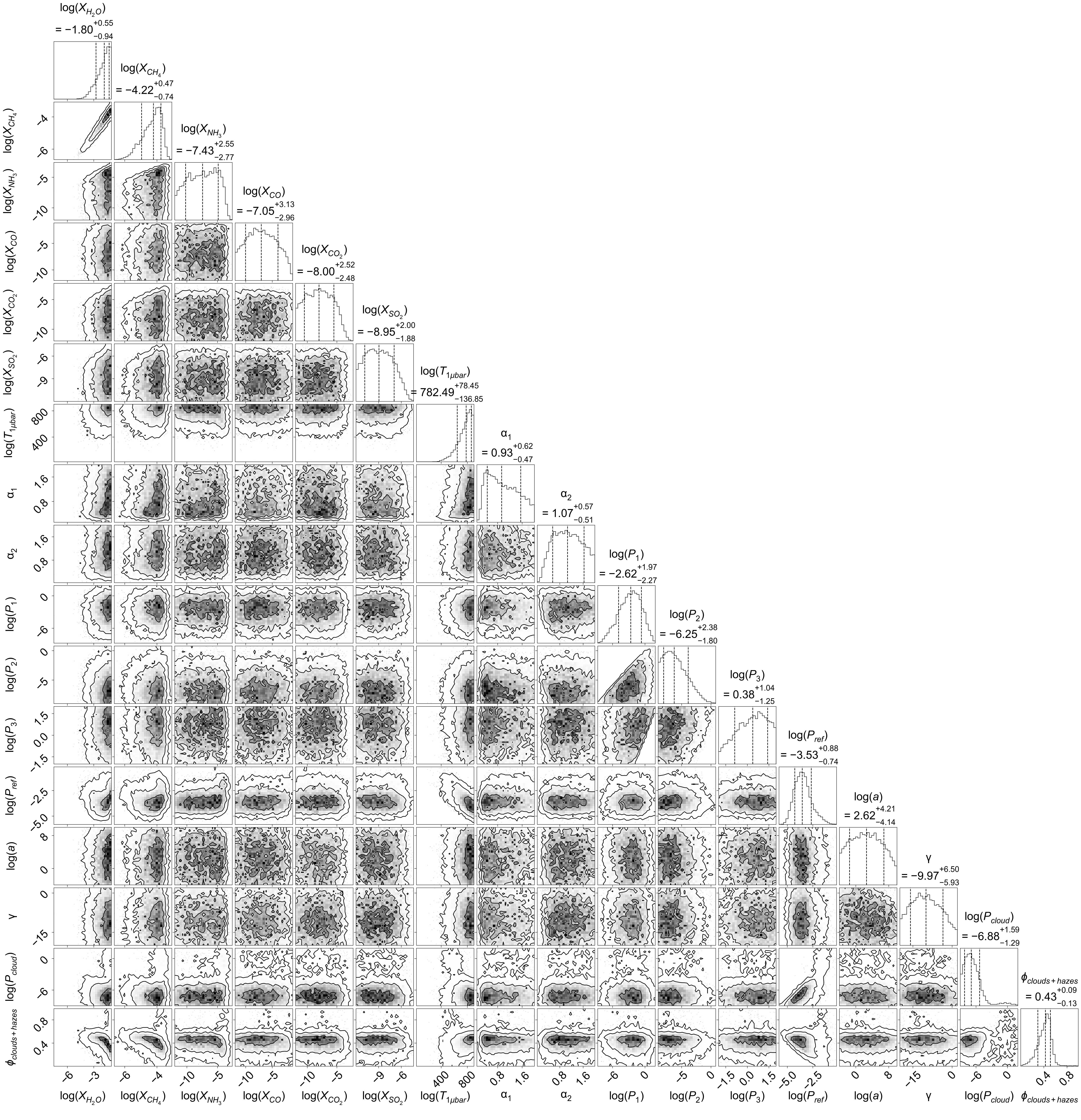}
    \caption{\textbf{Covariances in the free transmission retrieval.} Posterior distribution for the free retrieval of JWST NIRCAM F322W2 transmission spectra using Aurora.}
    \label{fig:aurora_transmission}
\end{figure*}

\begin{figure*}[!htbp]
    \centering
    \includegraphics[width=\linewidth]{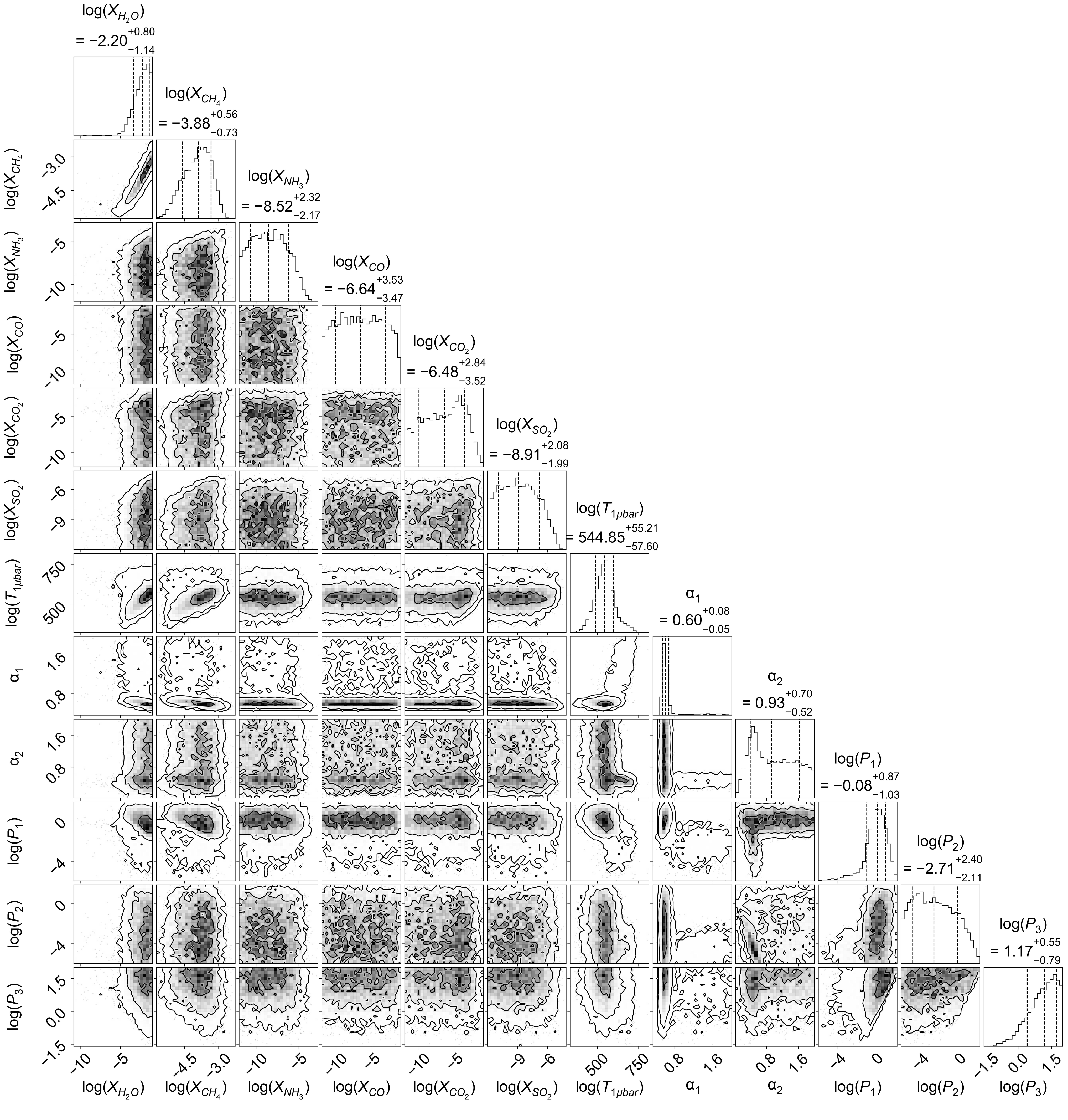}
    \caption{\textbf{Covariances in the free emission retrieval.} Posterior distribution for the free retrieval of JWST NIRCAM F322W2 emission spectra using Aurora.}
    \label{fig:aurora_emission}
\end{figure*}

\begin{figure*}
    \centering
    \includegraphics[width=\linewidth]{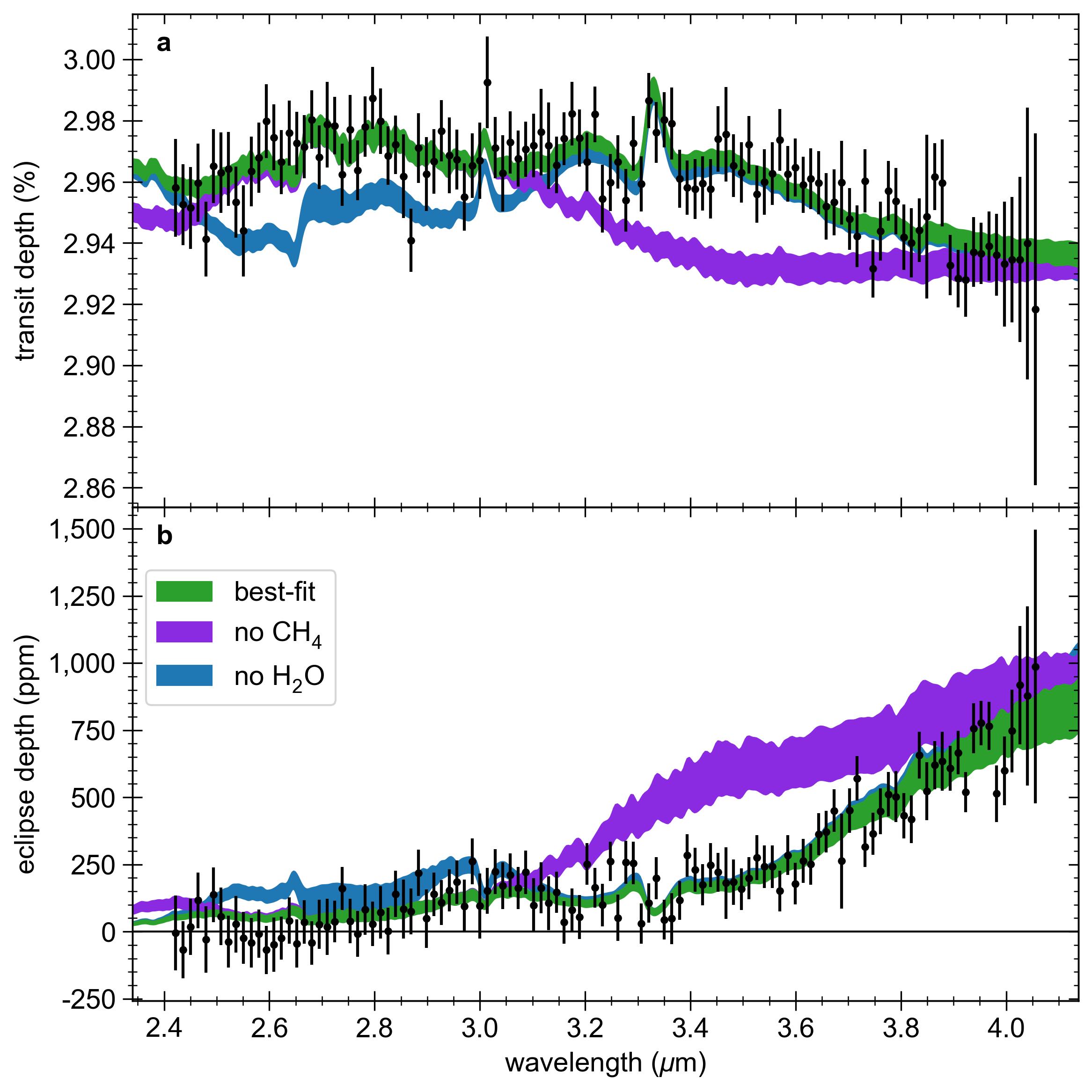}
    \caption{\textbf{Summary of the 1D-RCPE grid-based retrieval fits.} The observed transmission (\textbf{a}) and emission (\textbf{b}) spectrum (with 1$\sigma$ error bars) compared to an ensemble of from the 1D-RCPE retrieval fits, summarized with a 68\% confidence band derived from 200 posterior samples.  The contribution from the major absorbers are indicated by removing them (``no CH$_4$", ``no H$_2$O") from the ``best-fit" model during the posterior sampling spectral post-processing.}
    \label{fig:retrieval_grid}
\end{figure*}

\begin{figure*}[!htbp]
    \centering
    \includegraphics[width=\linewidth]{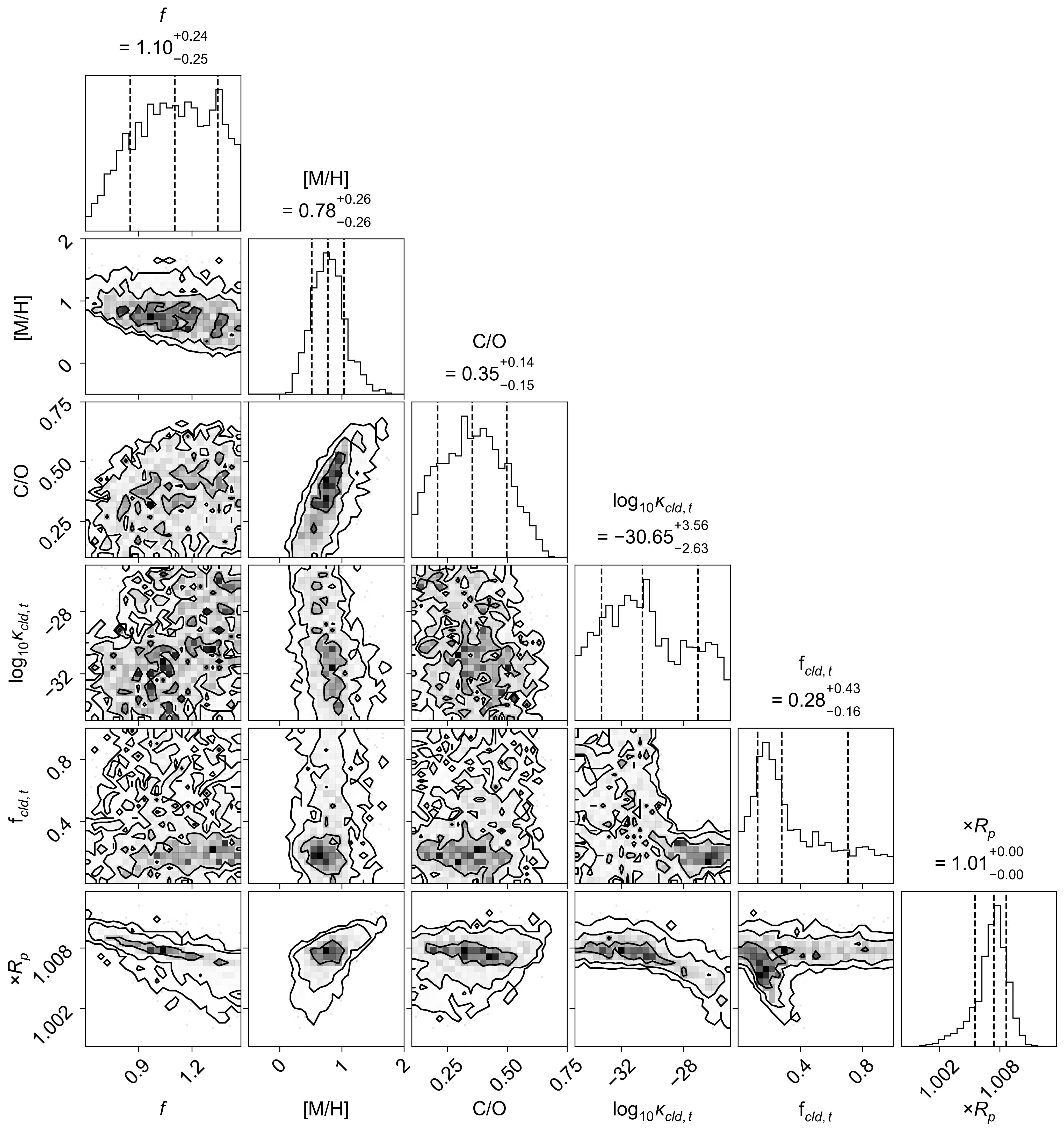}
    \caption{\textbf{Covariances in the 1D-RCPE transmission grid-based retrieval.} Posterior distribution for the 1D-RCPE grid-based retrieval of JWST NIRCAM F322W2 transmission spectra using ScCHIMERA.}
    \label{fig:1DRCPE_transmission}
\end{figure*}

\begin{figure*}[!htbp]
    \centering
    \includegraphics[width=0.7\linewidth]{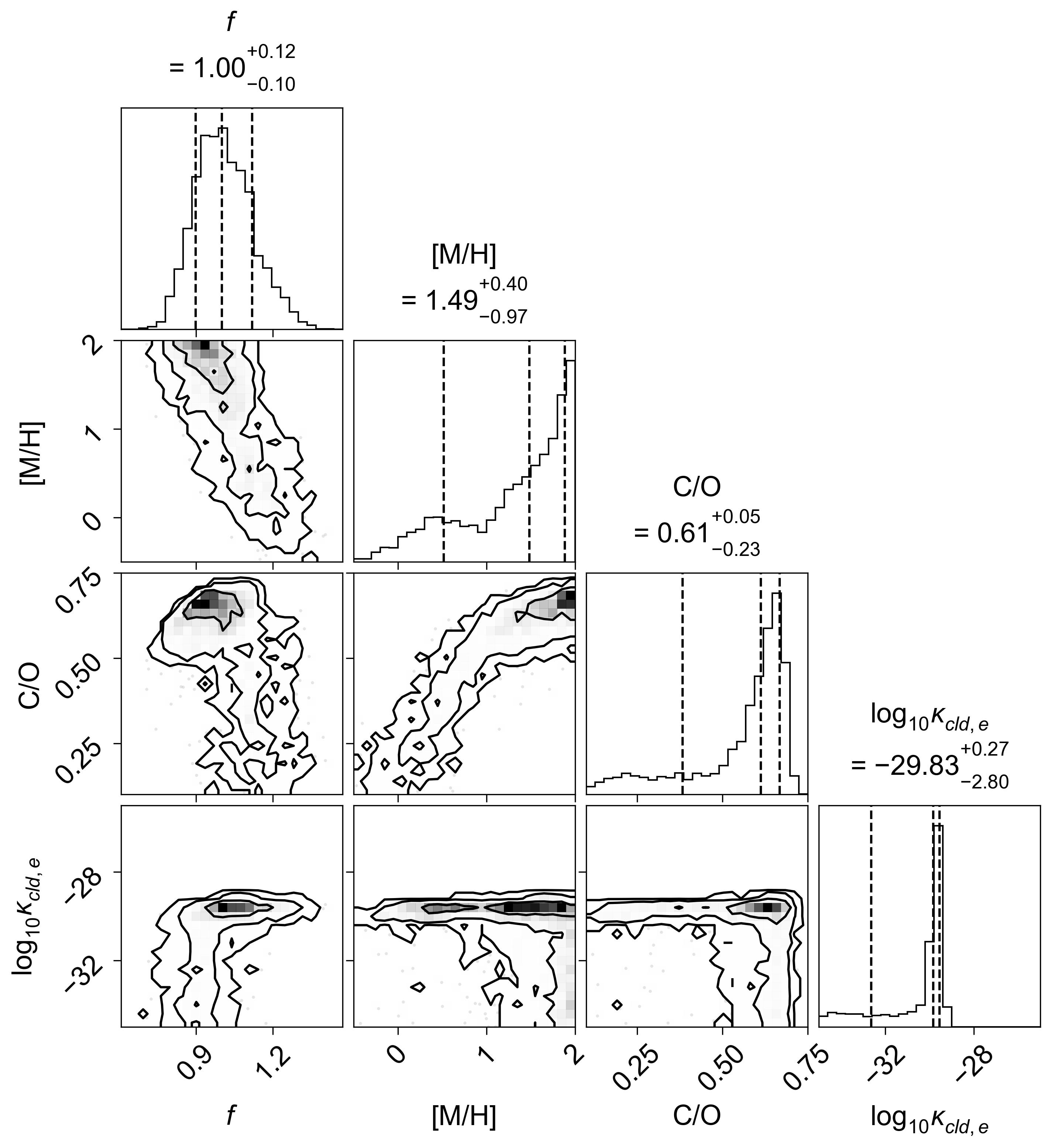}
    \caption{\textbf{Covariances in the 1D-RCPE emission grid-based retrieval.} Posterior distribution for the 1D-RCPE grid-based retrieval of JWST NIRCAM F322W2 emission spectra using ScCHIMERA.}
    \label{fig:1DRCPE_emission}
\end{figure*}

\end{appendices}
\end{document}